\documentclass[%
 notitlepage,
 reprint,
 superscriptaddress,
 amsmath,amssymb,
 aps,
]{revtex4-2}

\usepackage{xcolor}
\usepackage{lipsum}
\usepackage{booktabs}

\usepackage{graphicx}
\usepackage{dcolumn}
\usepackage{bm}
\usepackage{hyperref}

\usepackage{epstopdf}

\newcommand{\ket}[1]{\vert #1 \rangle}

\begin{document}

\preprint{APS/123-QED}

\title{Binary Apollonian networks}

\author{Eduardo M. K. Souza}
\affiliation{
Departamento de F\'isica, Universidade Federal de Sergipe, 49100-000 S\~{a}o Crist\'{o}v\~{a}o, Sergipe, Brazil
}%

\author{Guilherme M. A. Almeida}
\affiliation{
Instituto de F\'isica, Universidade Federal de Alagoas, 57072-900 Macei\' o, Alagoas, Brazil
}%

\date{\today}

\begin{abstract}
There is a well-known relationship between the binary Pascal's triangle and Sierpinski triangle in which the latter obtained from the former by successive modulo 2 additions on one of its corners. Inspired by that, we define a binary Apollonian network and obtain two structures
featuring a kind of dendritic growth. They are found to inherit the small-world and scale-free property from the original network but 
display no clustering.
Other key network properties are explored as well.  Our results reveal that
the structure contained in the Apollonian network may be employed to 
model an even wider class of real-world systems.
\end{abstract}


\maketitle

Network theory has brought significant advances to the field of complex systems,
being applied to the spread of diseases and information, author collaboration networks, engineering, condensed-matter physics, transport, and much more \cite{alb,newbw}. 
While many types of networks are
used to model the dynamics of those and other classes of systems, they often share
a lot in common. 
To bypass complexity and reach for perspective, it is thus valuable
to explore their universal properties.
For instance, real-world networks
are neither fully random nor fully regular but 
fit somewhere in between. As such, many of them display
features belonging to the class of small-world networks \cite{watts}. These are
graphs where it is often possible to make from one node to another
by taking a small number of steps. Formally,
it means the characteristic path length 
grows logarithmically with the number of nodes, while retaining 
a clustering coefficient well above that of a random graph.

A paradigmatic example of a graph displaying the small-world property is the Apollonian network (AN) \cite{andrade} (see Fig. \ref{fig2}), named after
Apollonius' circle-packing ideas. This class of networks is also scale-free (the degree distribution follows a power law) and can be embedded in Euclidean plane, which is quite relevant to condensed-matter physics.  ANs have since been
used in the context of strongly-correlated electron systems \cite{souza07}, Bose-Einstein condensation \cite{oliveira2}, localization \cite{cardoso}, quantum dynamics \cite{xu08,alme}, and other fields \cite{arca,andrade05,oliveira,andrade09}. In particular, its topology yields ubiquitous transport properties as the spectrum
is made up by discrete, localized modes as well as extended modes, with a high level of degeneracy due to the node degree distribution and
$2\pi/3$ rotational symmetry \cite{alme}. 

\begin{figure}
\includegraphics[width=0.35\textwidth]{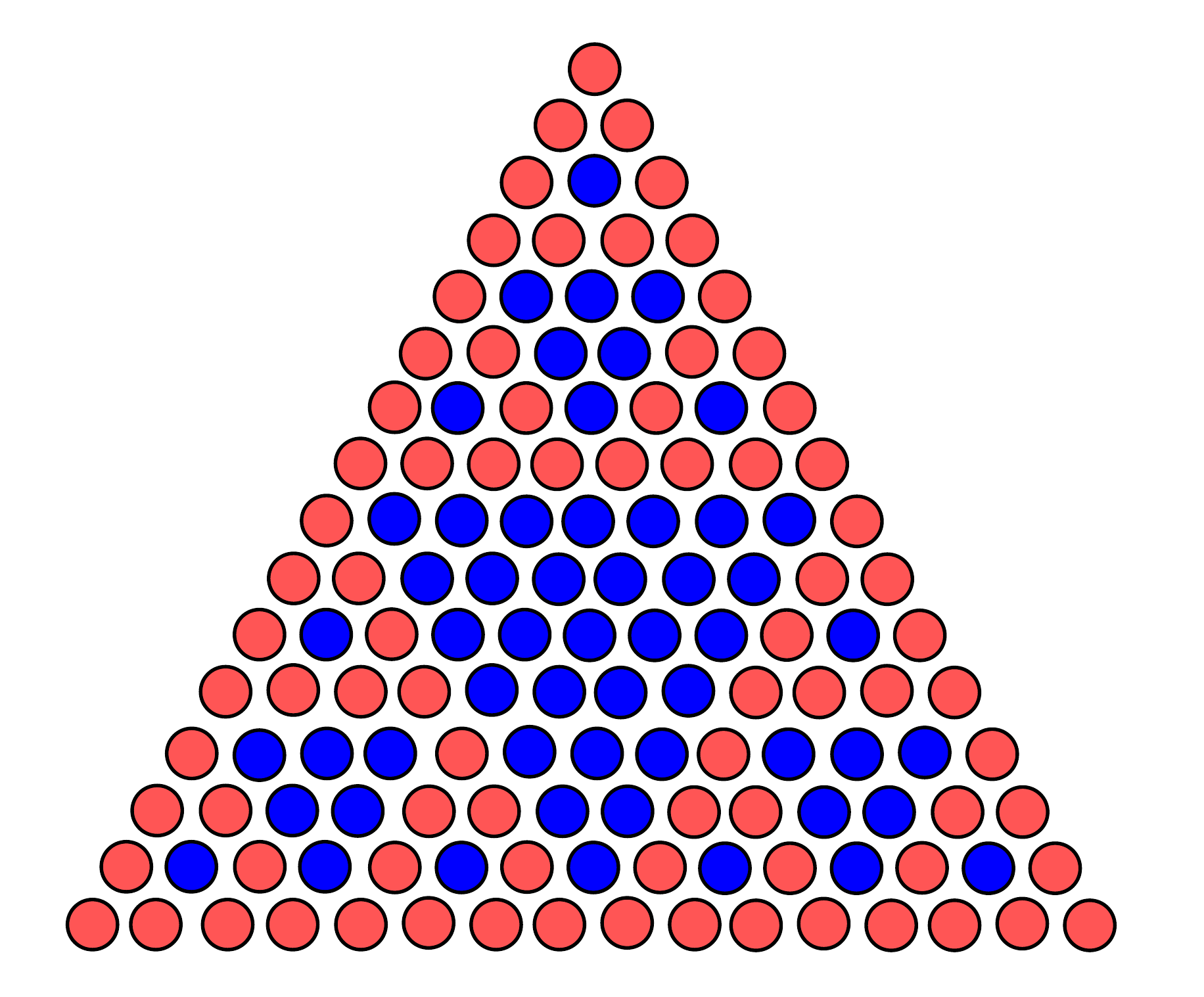}   
    \caption{A structure resembling the Sierpinski fractal triangle can be obtained from the famous Pascal's triangle by carrying out successive modulo 2 additions downward. Red (light) circles stand for bit 1 and blue (dark) circles represent bit 0.}
    \label{fig1}
\end{figure}

\begin{figure*}[t!]
\includegraphics[width=0.5\textwidth]{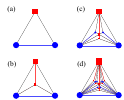}   
    \caption{Apollonian network from the (a) zeroth generation (n=0) through the (d) third (n=3). The total number of sites is $N_{n} = (3^{n}+5)/2$, and the number of edges is $B_{n} = 3N_{n}-6$. To build its binary version we assign a bit to each one of the nodes. Here, blue circles (red squares) represent bit 0 (1). Each node belonging to subsequent generations is set according to the modulo 2 sum of the three nodes surrounding it that form a triangle. The Apollonian subnetworks are defined by detaching both groups of bits keeping their own bonds.}
    \label{fig2}
\end{figure*}

The primary goal of investigating complex networks is
to provide insights into the behavior of real-world systems, either natural or manufactured. In order to do that, one 
should be able to identify them accurately, something not always taken for granted. On the other hand, 
seemingly distinct structures may display very similar behavior. From a more fundamental level, we may ask ourselves what are the main ingredients responsible for that. Following this direction, in this work we set out to take a better look into the building blocks of the
AN. We derive subnetworks inspired by the well known relationship between the binary Pascal's triangle and the Sierpinski triangle.
The latter is a fascinating fractal structure that emerges from various systems in nature and 
is connected to many areas of mathematics. It can be obtained, for example, from the Pascal's triangle by 
assigning a binary number, 0 or 1, to each element based on its even or odd parity, respectively. 
It is easy to see that the larger the triangle is (see Fig. \ref{fig1}), its binary structure converges to the Sierpinski fractal. 
It can also be generated from scratch by performing successive modulo 2 additions in the Pascal's triangle. 

Based on such construction we define a binary AN where 
to each node is assigned a bit that is the result of the modulo 2 addition of
the three nodes (from the previous generation) surrounding it. Two subnetworks arise from such procedure and we will address them
in detail by evaluating standard measures such as the degree distribution, clustering coefficient, mean distance, and spectrum 
of the adjacency matrix.



The regular AN is built starting out with a triangle (zeroth generation; $n=0$) as depicted in Fig. \ref{fig2}. 
In the following generation ($n=1$), a new node is
added within it and the edges now form three triangles. 
Subsequent generations are obtained by repeating this process for each triangle to render a self-similar network.

\begin{figure}[t]
	\includegraphics[width=0.35\textwidth]{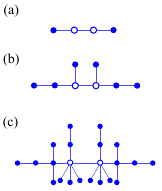}   
	\caption{Apollonian subnetwork made up by the bit-0 nodes (0-AsN) for generations (a) $n=2$ (b) $n=3$, and (c) $n=4$. Empty circles represent both nodes located at the bottom corners of the original network. }
	\label{fig3}
\end{figure}

\begin{figure}[t]
	\includegraphics[width=0.30\textwidth]{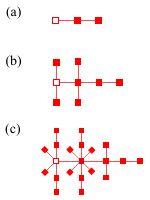}   
	\caption{Apollonian subnetwork made up by the bit-1 nodes (1-AsN) for generations (a) $n=2$ (b) $n=3$, and (c) $n=4$. Empty square represent the node located at the top corner of the original network. }
	\label{fig4}
\end{figure}

To build a binary version of the AN
we assign a bit to each node. Consider the blue circles (red squares) represent bit 0 (1) in Fig. \ref{fig2}.
Subsequent nodes are defined according
to the modulo 2 addition of the bits that form their respective triangle.
From generation $n=0$ we note 
that out of the many possibilities to set the three initial bits, they will either deliver 
the standard network or amount to the construction seen in Fig. \ref{fig2}. This summarizes the output
whenever one bit is different from the other two. 

We have obtained two groups of nodes in the AN.
If we now split them up keeping only their own links (edges connecting different bits are destroyed), a couple of subnetworks 
emerge.
Figures \ref{fig3} and \ref{fig4} shows each one of them in detail from a clearer perspective starting from $n=2$. For convenience, we will
refer to those as 0-AsN and 1-AsN, respectively, from now on. 
Both structures can be easily treated by their own
growth rules without looking back at the original AN. 
For each new generation, every node is joined by a number of new others equal to
the number of edges it had in the previous generation, with few exceptions. In the 0-AsN,
the nodes located at the bottom corners of the original network (see Fig. \ref{fig2}), represented by the
empty circles in Fig. \ref{fig3}, acquire one bond less when
going from an even to odd generation.
In the 1-AsN, the node located
at the top corner of the AN, tagged as the empty square in Fig. \ref{fig4}, 
gets an extra neighbor when going from an even to odd generation and one less otherwise.
Note that the growth process of such networks resembles those
seen in deposition of materials that develop a dendritic pattern \cite{bara2}.

The total number of sites in the standard, $n-$th generation AN is $N_{n} = (3^{n}+5)/2$ and the number
of edges $B_{n} = 3N_{n}-6$ \cite{andrade}. 
For the 0-AsN and 1-AsN, it is straightforward to obtain that 
$N_{n} =(3^{n}+(-1)^{n}+6)/4 $ and $N_{n} = (3^{n}-(-1)^{n}+4)/4$, respectively. The number of edges is
$B_{n} =N_{n}-1$ for both cases.
 
 
We now move on to evaluate some fundamental properties of both subnetworks and compare them 
with the original AN \cite{andrade}. 
Let us start with the degree of a node, that is the number of edges connected to it, 
$k_i = \sum_{j} A_{i,j}$, where $A_{i,j}$ are the entries of the corresponding adjacency matrix $\mathrm{\mathbf{A}}$ ($A_{i,j} = 1$ 
if there is an edge between nodes $i$ and $j$, with $A_{i,j} = 0$ otherwise).
More precisely, in the AN there are 
$m(k,n)=3^{n-1}$, $3^{n-2}$, $3^{n-3}, \ldots , 3^{2}$, $3$, $1$, and $3$
sites with degree
 $k=3$, $3 \times 2$, $3 \times 2^{2}, \ldots , 3 \times 2^{n-2}$, $3 \times 2^{n-1}$, and $2^{n} + 1$, respectively, at the $n-$th generation. 
Similarly, we find that the 0-AsN features (for $n>2$)
$m(k,n)=(3^{n-1}+(-1)^{n})/2,\ldots, 4, 2, 2$ sites with degree 
$k=2^{0}, \ldots,2^{n-3}$ ,$2^{n-2}, (2^{n+1}+(-1)^{n}+3)/6$. 
The 1-AsN has  (for $n>2$)
$m(k,n)=(3^{n-1}+(-1)^{n-1})/2, \ldots ,1,1,1$ sites with degree
$k=2^{0}, \ldots , 2^{n-2}, (2^{n}+(-1)^{n-1})/3, 2^{n-1}$.

\begin{figure}
	\includegraphics[width=0.35\textwidth]{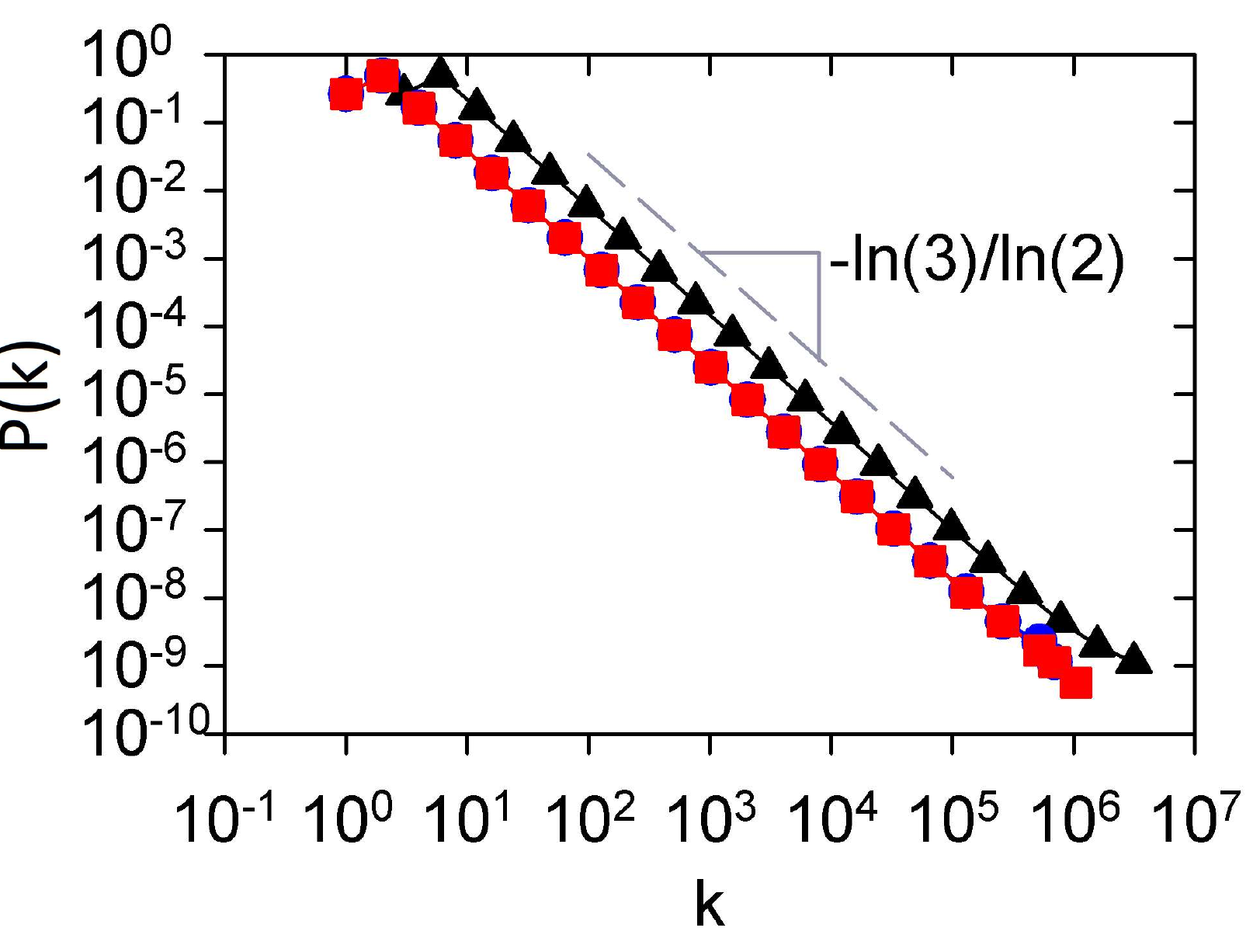}   
	\caption{Degree distribution $P(k)$ for the
		AN (black triangles), 0-AsN (blue circles) and 1-AsN (red squares). All of them display the scale-free property, $P(k) \propto k^{-\gamma}$, with $\gamma = \ln3/\ln2$.}
	\label{fig5}
\end{figure}

Using the information above we can show the subnetworks are scale-free, just like the AN.  This property can be 
confirmed through the cumulative distribution
\begin{equation} \label{eq:eqpk}
P(k) = \sum_{k' \ge k} \frac{m(k',n)}{N_n},
\end{equation}
which is displayed in Fig. \ref{fig5}.
For $n \gg 1$, we get the typical asymptotic power-law decay $P(k) \propto k^{-\gamma}$. The  
exponent can be evaluated directly from the expressions for $m(k,n)$, $N_{n}$, resulting in 
$\gamma = \ln3/\ln2=1.585$ for all the networks.
Note that such value is exactly the Hausdorff dimension of the Sierpinski triangle \cite{fal}, which is 
not a mere coincidence, since it has the same combinatorial structure as the Apollonian gasket.

\begin{figure}
	\includegraphics[width=0.35\textwidth]{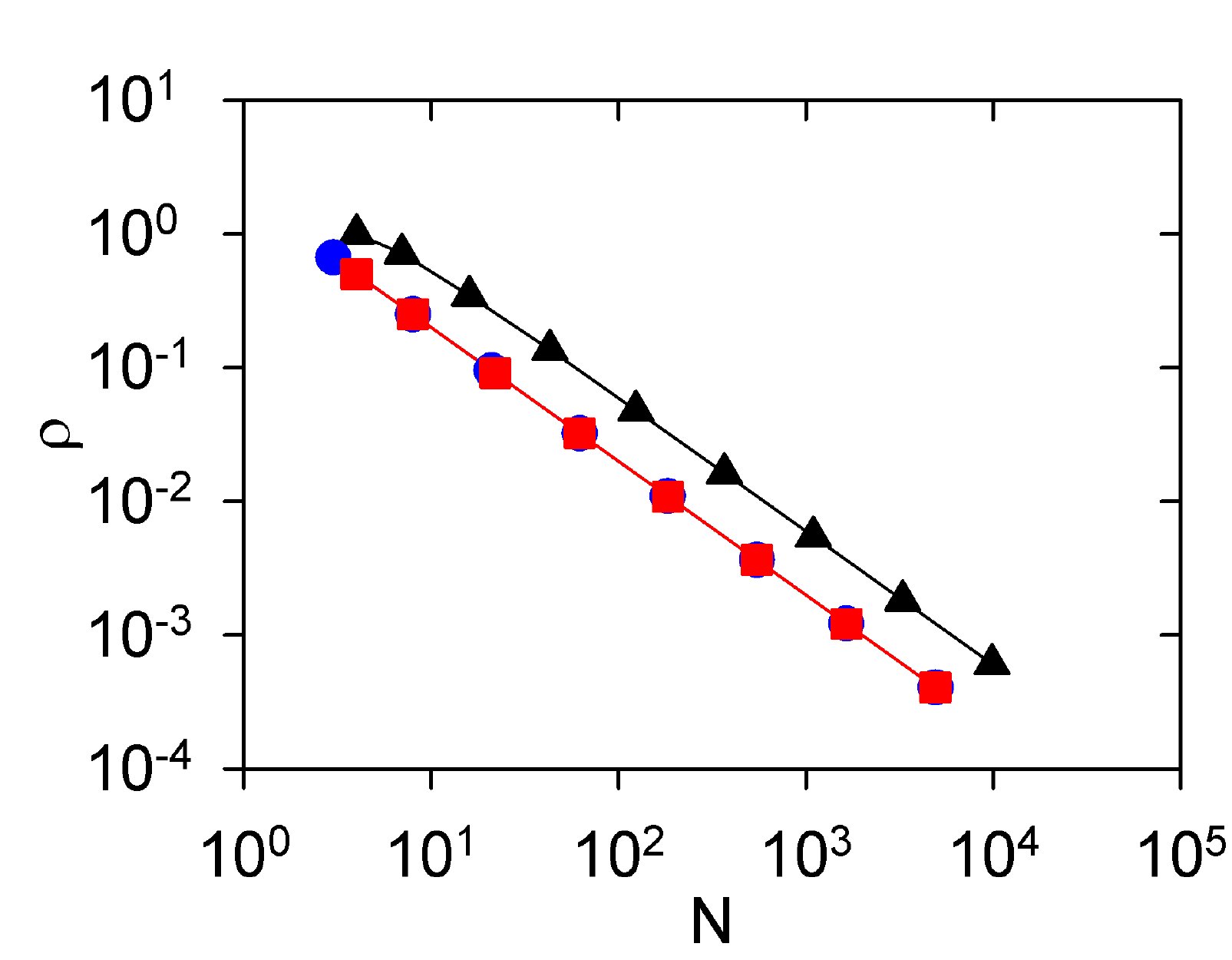}   
	\caption{Connectance $\rho = \sum_{i,j} A_{i,j} /N(N-1)$ versus network size $N$ for the
		AN (black triangles), 0-AsN (blue circles) and 1-AsN (red squares). In each of these $\rho \propto N^{-1}$.}
	\label{fig6}
\end{figure}

Let us now check their connectance, defined by
$\rho = 2B_n/N(N-1) = \sum_{i,j} A_{i,j} /N(N-1)$, where $N(N-1)/2$ is 
the maximum number of edges for a $N$-site network. Note that $\rho \in [0,1]$ and basically quantifies how 
sparse a network is.
Figure \ref{fig6} shows such measure for the three networks. They all yield $\rho \propto N^{-\alpha}$, with $\alpha=1$, thus becoming increasingly sparse with $N$. 
At that regime the average degree tends to a constant, namely $\langle
k \rangle = N_{n}/B_{n}=1-1/N_{n} \rightarrow 1$ (0-AsN and 1-AsN) and $\langle
k \rangle = 3-6/N_{n} \rightarrow 3$ (AN).  

\begin{figure}
	\includegraphics[width=0.35\textwidth]{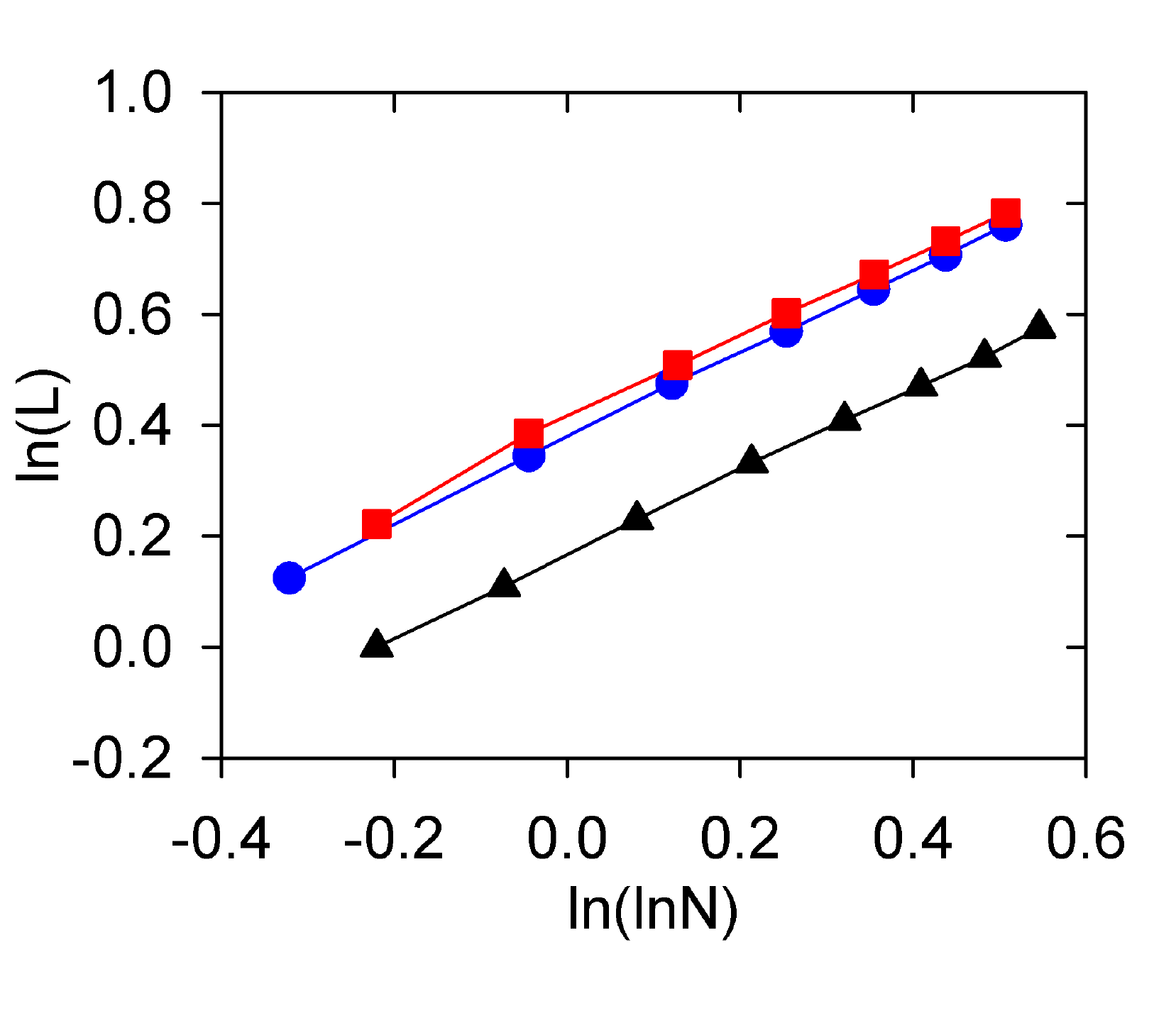}   
	\caption{Log of average length
of the shortest path versus $\ln (\ln N)$ for the
		AN (black triangles), 0-AsN (blue circles) and 1-AsN (red squares). Each gives $L \propto (\ln N)^{\beta}$,
with $\beta=3/4$.}
	\label{fig7}
\end{figure}

The AN is specially known for its small-world property. 
To account for this the average length
of the shortest path is usually employed:
\begin{equation} \label{eq:lm}
L = \frac{1}{N(N-1)}\sum_{i \neq j} d_{i,j},
\end{equation}
where $d_{i,j}$ is the length of the shortest path between nodes $i$ and $j$. 
In Fig. \ref{fig7} we evaluate it for both subnetworks, the AN, and 
note that
$L \propto (\ln N)^{\beta}$,
with $\beta=3/4$ signalling the intermediate behavior 
between small- ($L \propto \ln N$) and ultra-small-world ($L \propto \ln(\ln N)$) networks found in \cite{andrade} for the AN.
Now, the network diameter $d_{n}=L_{max}$, that is the maximum shortest path among all the pairs of sites, 
is $d_{n}=2n/3$ for the AN \cite{a3d}, considering $n \gg 1$ and $d_{n}=2n+1$ ($d_{n}=2n$) for the 0-AsN (1-AsN).
Hence, for large $N$ we have that $L_{max} \approx 2\log(N)/(3\log(3))$ in the AN and $L_{max} \approx 2\log(N)/\log(3)$ in both subnetworks.

One striking difference between the AN and the subnetworks comes from the local clustering coefficient $C$.
It is known that a high value of $C$ is another signature of the small-world property.
However, while the in AN it tends to $C=0.828$ for large $N$ \cite{andrade}, the subnetworks yield $C=0$.
As mentioned before, they are tree-like structures and given $C$ is proportional to the number of triangles in a network, one readily concludes that $C=0$.
Nevertheless, their small-world character is due to the presence of hubs (see Figs. \ref{fig3} and \ref{fig4}). 

\begin{figure}
	\includegraphics[width=0.35\textwidth]{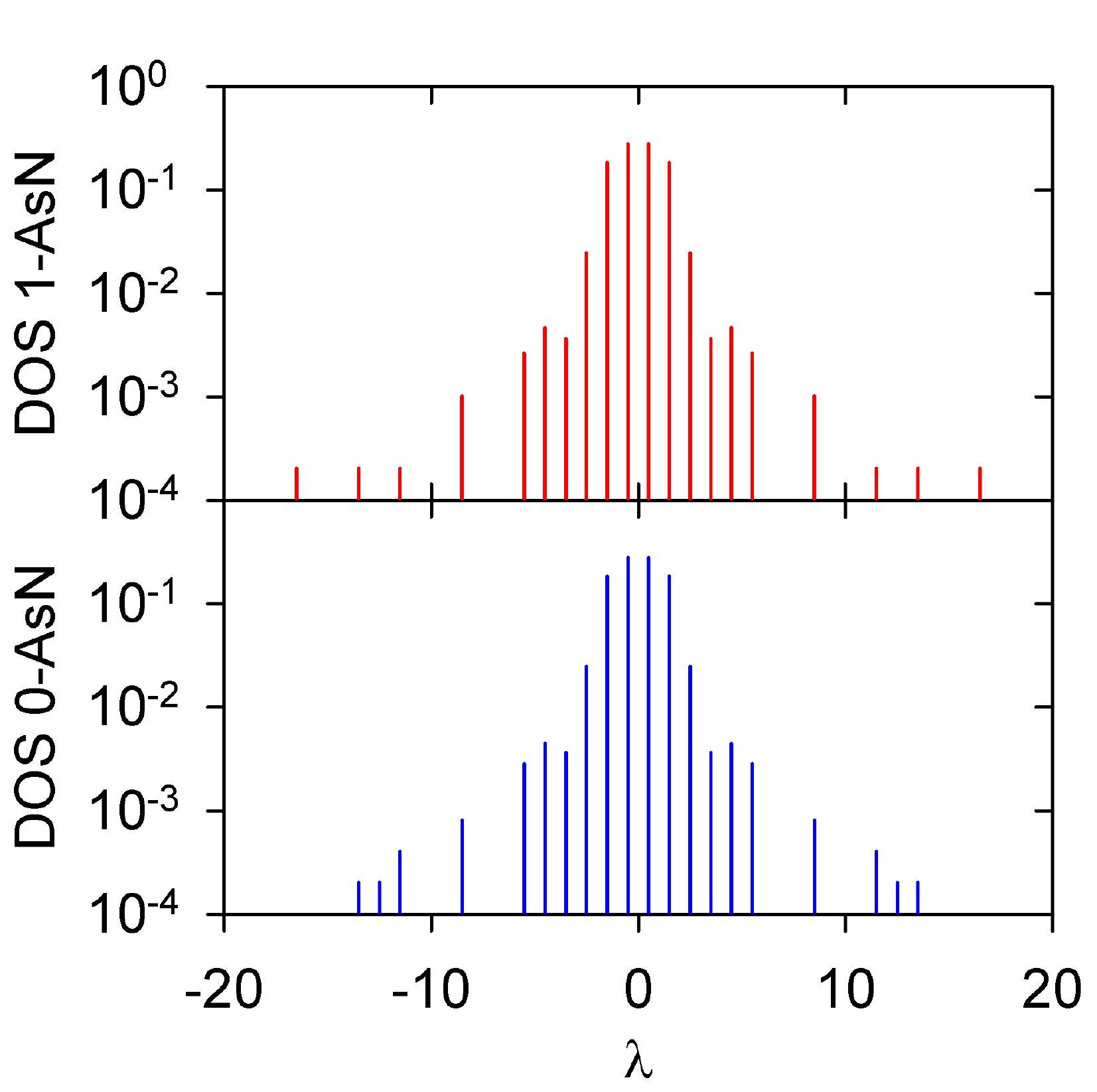}   
	\caption{Density of states of the adjacency matrix of the 0-AsN and 1-AsN for generation $n=9$.}
	\label{fig8}
\end{figure}

\begin{figure}
	\includegraphics[width=0.35\textwidth]{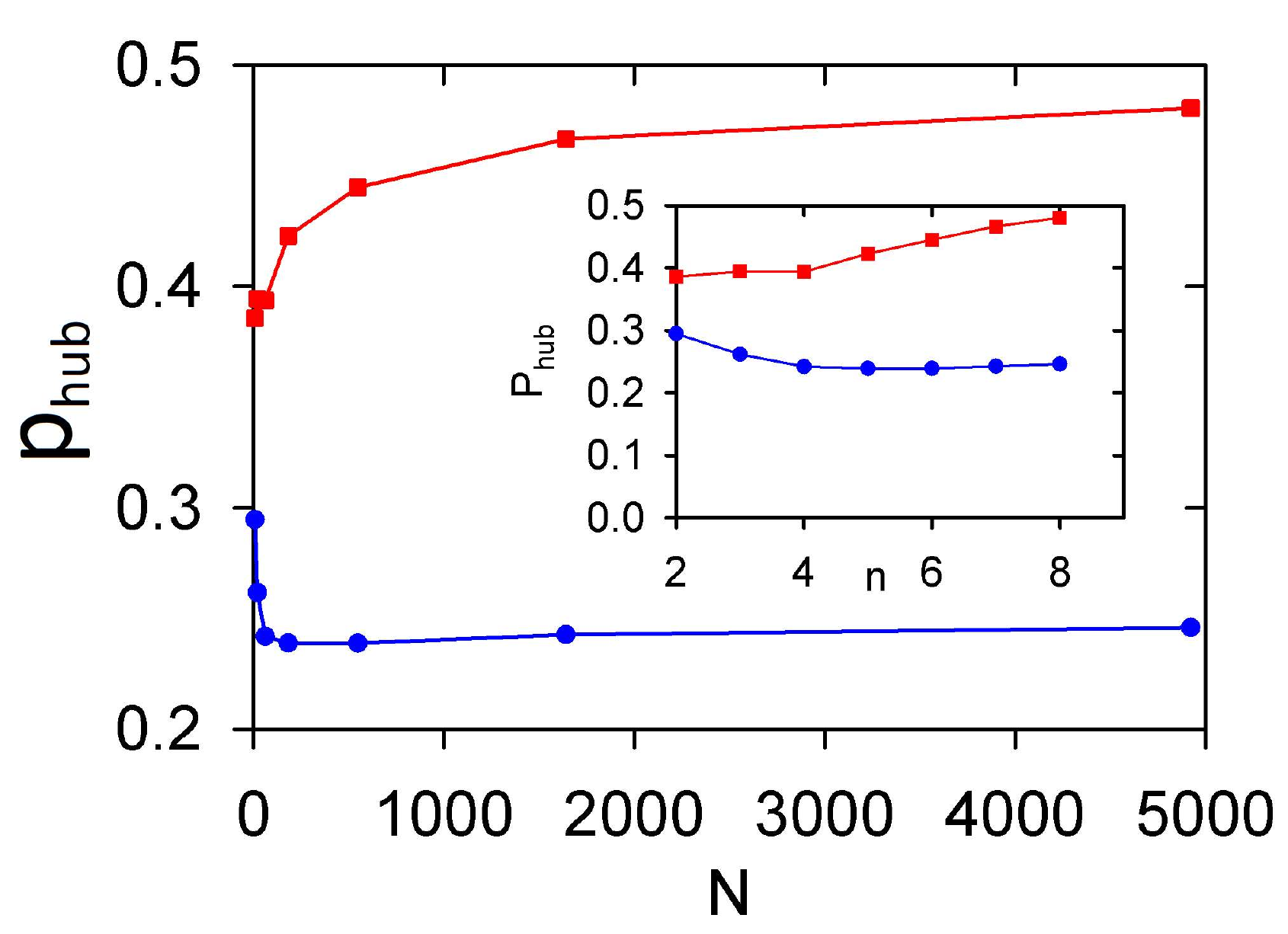}   
	\caption{Probability amplitude of the hub node $p_{\mathrm{hub}}$ versus $N$ (and generation $n$; see inset) considering the eigenvector having largest eigenvalue. The 0-AsN (1-AsN) is depicted by blue circles (red squares).}
	\label{fig9}
\end{figure}

\begin{figure}
	\includegraphics[width=0.35\textwidth]{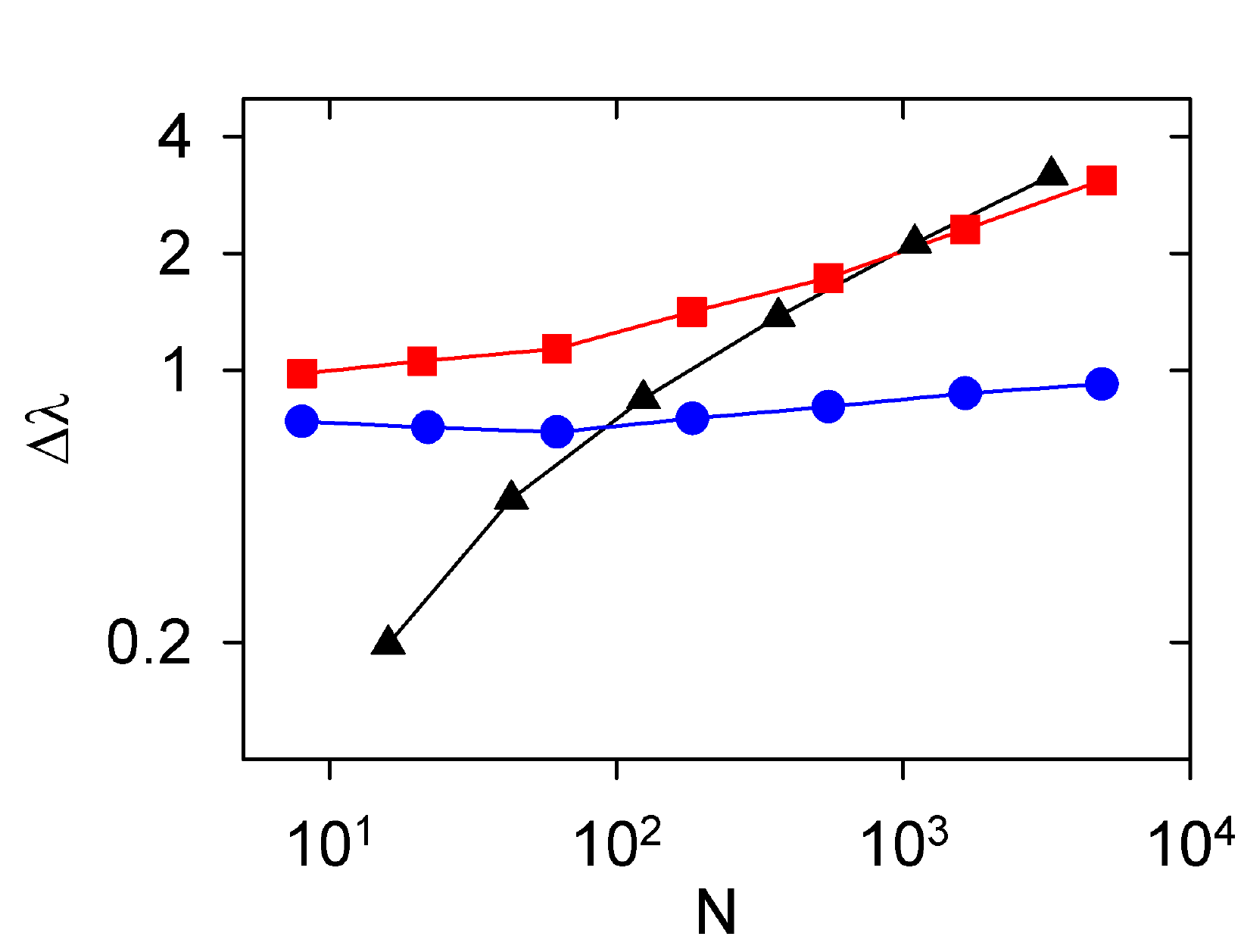}   
	\caption{Gap between the ground and first excited state $\Delta \lambda$ for the AN (black triangles), 0-AsN (blue circles), and 1-AsN (red squares).}
	\label{fig10}
\end{figure}

We now move on to study some spectral properties of the adjacency matrix $\mathrm{\mathbf{A}}$.
Note that both subnetworks have unidirectional edges. That is to say, their adjacency matrices are symmetric and therefore have spectral decomposition.
They are also bipartite networks, meaning that the nodes can be split into two disjoint sets, say $S_1$ and $S_2$, in such a way the nodes belonging to $S_{1}$ ($S_{2}$) only share direct links with those in $S_{2}$ ($S_{1}$).
As a consequence, the eigenvalue spectrum $\lbrace \lambda_i \rbrace$ is symmetric,  $\lambda_{N+1-i} = -\lambda _{i}$ for all $i$. 
In Fig. \ref{fig8} we show the density of states $DOS(\lambda)=\frac{1}{N}\sum_i \delta ( \lambda-\lambda_{i} )$, $\delta$ being Dirac delta function, for the ninth generation.
In both the 0-AsN and 1-AsN most of the states populate the center of the band. One clear difference is that the 1-AsN 
is able to span more extreme modes. 
In the regular AN, modes lying far away from the center of the band carry a high degree of localization, mostly due to the nodes having highest number of connections \cite{cardoso, alme}. To see how it goes for the subnetworks
let us analyze the eigenvector with largest eigenvalue $\ket{\lambda_{\mathrm{max}}}=[v_1,v_2,\ldots,v_N]^T$, with $p_{i}=|v_{i}|^2$ 
being the probability amplitude for node $i$. 
Figure \ref{fig9} shows such amplitude for the nodes having the maximum degree against size $N$. 
Note from Figs. \ref{fig3} and \ref{fig4} that the 0-AsN contains two equivalent hubs whereas the 1-AsN holds
the hub of the original AN (cf. Fig. \ref{fig2}). 
We observe that $p_{\mathrm{hub}}$ converges to $1/4$ ($1/2$) as $N$ grows on the 0-AsN (1-AsN). 
In \cite{souza} it was analytically proved that for any bipartite lattice  $\sum_{i \in S_{j} } p_{i}$ is independent of $\lambda$. 
Here, both subnetworks fulfill $\sum_{i \in S_{1} } p_{i}= \sum_{i \in S_{2} } p_{i}=1/2$ and therefore, for large $N$,
the eigenvector tends to become fully localized at the hub.

Last but now least, Fig. \ref{fig10} displays the energy gap $\Delta \lambda$ between the ground and the first excited state.
In the case of large $N$ the three networks obeys a power-law scaling $\Delta \lambda \propto N^{\theta}$, with $\theta = 0.34$
for the AN and $\theta =0.06 (0.26) $ for the 0-AsN (1-AsN).
This implies the connectivity of the nodes from younger generations diverge as $N$ increases \cite{las,goh}.

\begin{table}[t!]
\caption{Main properties of the Apollonian subnetworks in comparison with the original AN. $\langle k \rangle$ is the average degree, $C_{N\rightarrow \infty}$ is the clustering coefficient, $d_{n}$ is the network
diameter, $\alpha$ tells how the connectance $\rho \propto N^{-\alpha}$, $\beta$ is the small-world exponent, $\gamma$ is the degree
distribution exponent, and $\theta$ comes from the power-law scaling for the gap between ground and first excited states $\Delta \lambda \propto N^{\theta}$.}
\label{table}
\begin{tabular}{@{}lllll@{}}
\toprule
 &  AN &  0-AsN & 1-AsN \\ \midrule
 $\langle k \rangle$ & $3-\frac{6}{N_{n}}$ & $1-\frac{1}{N_{n}}$ & $1-\frac{1}{N_{n}}$ \\ 
	$C_{n \rightarrow \infty}$& $ 0.828$ & $0$ & $0$ \\ 	
	$d_{n}$ & $x$ & $2n+1$ & $2n$ \\ 
	$\alpha$& $1.0$ & $1.0$ & $1.0$ \\ 
	$\beta$& $3/4$ & $3/4$ & $3/4$ \\ 
	$\gamma$ & $\ln(3/2)$ & $\ln(3/2)$ & $\ln(3/2)$ \\ 
	$\theta $& $0.34$ & $0.06$ & $0.26$ \\  \bottomrule
\end{tabular}
\end{table}


We have obtained two subnetworks from the standard AN by assigning a bit 
to each one of the nodes and 
setting its growth to follow successive
modulo 2 additions on a binary version of it. 
Those novel structures happen to feature their own growth rules, 
developing a dendritic pattern. 
We found that both subnetworks share many characteristics with their embedded AN, including 
the small-world property. However, the clustering coefficient is zero due to the fact that no triangles
can be formed among the nodes.
Table \ref{table} summarizes our main findings in comparison with the original AN. 

Our findings reveal that many emerging structures found in nature as well as in manufactured systems, such 
those that feature multi-branch tree growth, may share more similarities with the deterministic AN \cite{andrade} than one would think.

\section{Acknowledgments}
We thank A. M. C. Souza for presenting us with the original idea and sharing key insights. 
This work was supported by CNPq.

\end{document}